\documentclass[10pt]{iopart}
\usepackage{graphicx,rotating,subfigure,iopams} 
\usepackage{umlaut}
\usepackage{color}

\usepackage{graphicx}
\usepackage{dcolumn}
\usepackage{bm}
\usepackage{epsfig}
\renewcommand{\vec}[1]{\boldsymbol #1}
\newcommand{\be}{\begin{eqnarray}}
\newcommand{\ee}{\end{eqnarray}}

\def\refeq#1{(\ref{#1})}
\def\d{\mbox d}

\def\nn{\nonumber}
\def\i{\int_{-\infty}^{\infty}}

\def\Or{\mathcal O}

\def\al{\alpha}

\def\l{\left}
\def\r{\right}
\def\te{\mbox{e}}
\def\rmi{{\rm i}}

\def\up{\uparrow}
\def\Up{\Uparrow}
\def\down{\downarrow}
\def\Down{\Downarrow}
\def\tr{\mbox{tr}}

\begin{document}
\bibliographystyle{jpa}
\title[Spin-dynamics in the central spin model]{Spin- and entanglement-dynamics in the central-spin model with homogeneous couplings}
\author{Michael Bortz}
\address{Department of Theoretical Physics, Research School of Physics and Engineering, Australian National University, Canberra ACT 0200, Australia}
\author{Joachim Stolze}
\address{Institut f\"ur Physik, Universit\"at Dortmund, 44221 Dortmund, Germany}
\begin{abstract}
We calculate exactly the time-dependent reduced density matrix for the central spin in the central-spin model with homogeneous Heisenberg couplings. Therefrom, the dynamics and the entanglement entropy of the central spin are obtained. 
A rich variety of behaviors is found, depending on the initial 
{state} of the bath spins.
For an initially unpolarized unentangled bath, the polarization of the central spin
decays to zero in the thermodynamic limit, 
while its
entanglement entropy becomes maximal. On the other hand, {if the unpolarized environment is initially in an eigenstate {of the total bath spin}, the central spin and the entanglement entropy exhibit persistent monochromatic large-amplitude oscillations. This raises the question {to what extent}
entanglement of the bath spins prevents decoherence of the central spin.} 
\end{abstract}
\section{Introduction}
The central-spin model, or Gaudin model, describes one spin coupled to $N-1$  bath spins via  isotropic Heisenberg interactions,  
\be
H=\sum_{j=1}^{N-1}A_j \vec S_0 \cdot \vec S_j + h S_0^z\label{gaudef}\,,
\ee
including a magnetic field which couples to the central spin only. Our focus here is on spin-1/2 particles. 

The importance of this model is manifold. From a quantum information processing point of view, it describes the interaction of one qubit with an environment, serving as a model to study decoherence processes. In this regard, it is used to capture the hyperfine interaction of an electron trapped in a quantum dot with the nuclear spins \cite{sch02,sch03,kha02,kha03,dob03,coi04,has06,den06,coi06}. In \cite{kha02,kha03}, it was found that {non-uniformities} in the couplings $A_j$ induce a non-exponential decay of the central spin. This result was obtained by a perturbative calculation for zero polarization of the bath spins, and by an exact approach in the thermodynamic limit for a fully polarized bath. 

Apart from that, from a theoretical point of view,  the challenging feature of this model is the following: The model is exactly solvable by Bethe ansatz (BA)\cite{gaubook}. On the other hand, in the thermodynamic limit,  exact results can be found {in} a mean-field approach \cite{yuz05,coi06}, where quantum mechanical operators are replaced by their expectation values. Especially, the compatibility of the quantum-mechanical solution for a large number of particles with the mean-field solution was addressed very recently \cite{coi06}. {There} the dynamical evolution of the central spin in the quantum and mean-field solutions was observed to be similar for zero magnetic field and homogeneous couplings. That observation was made with the bath being initially in a spin-coherent state with a fixed total spin.
The work summarized in the present article {shows explicitly that such a similarity between the classical and quantum solutions is not possible if the bath is initially in a product state} without entanglement between the bath spins.

Despite  the exact solvability of the model, far away from the thermodynamic limit, no systematic approach for dealing with the 
entire spectrum of the Gaudin model is known. Richardson \cite{ric77} gave a method to calculate finite-size corrections for the pairing Hamiltonian, whose BA equations are equivalent to those of the Gaudin model. Explicit results for finite-size corrections to the ground state and to low-lying excited states for the pairing Hamiltonian have been obtained \cite{yuz05b} starting from Richardson's work. However, these techniques have not been applied to the spectrum of the Gaudin model. The only resort seems to be numerics \cite{sch02,sch03,dom06,rom02}, which is hard enough a problem in itself. 

Given the importance of the Gaudin model from both theoretical and applied points of view it is highly desirable to obtain analytical results for an arbitrary number of particles $N$ with an arbitrary number of flipped spins $M$ (as compared to the ferromagnetic ``all up'' state) which corresponds to a total magnetization $S^z=N/2-M$. In this work, we consider the special case {of homogeneous couplings and zero external field,}
 $A_j\equiv 2 \forall j$ and $h=0$ (in \cite{kha03}, this model was considered for $M=1,2$). For this choice of parameters, the time-dependent reduced density matrix $\rho_0(t)$ of the central spin is calculated. Therefrom, the quantum mechanical average $\langle S_0^z\rangle(t)=\tr_0 \l\{S_0^z\rho_0(t)\r\}$ and the entanglement entropy $E(t)=-\tr_0\l\{\rho_0(t) \log_2\rho_0(t)\r\}$ are derived. {In all cases considered here, the central spin and the bath are uncorrelated initially, with the central spin in the ``down''-state $|\!\Down\rangle$, and the bath state specified further below.} 

Although it seems as if real physical systems like quantum dots are not described by homogeneous couplings \cite{sch03}, the study of that simple special case 
turns out to be very fruitful. First of all {it} is solvable explicitly for arbitrary $N,M$, which enables us to present exact results 
for finite particle number and arbitrary polarization. Indeed, highly non-trivial results, such as the decoherence time or the amount of screening of the central spin, could be obtained and are presented below. Furthermore, it is important to study the influence of different factors on the dynamics separately in order to 
assess their respective importance. These factors include the limit of large particle number, the magnetization and entanglement of the bath, the non-uniformity of the couplings and the magnetic field. The focus in this work is on the first three 
of these factors, 
with the couplings homogeneous and without a field. Finally, the study presented here is expected to be helpful for the 
investigation of generalizations (namely, to non-uniform couplings and finite $h$-field). Exact results for 
these more general settings are highly desirable from an applications point of view, and the study of the homogeneous case will serve as a point of reference.

Our main results are summarized in the following (units are chosen such that $\hbar\equiv 1$). {The simple structure of the Hamiltonian leads to an eigenvalue spectrum in which all transition frequencies (energy differences) are commensurate. Therefore all physical quantities are strictly periodic with period $\tau_P$, the Poincar\'{e} recurrence time. The recurrence time does not depend on the system size; $\tau_P=\pi$  $(2\pi)$ for $N$ even (odd). The detailed dynamics within every period $\tau_P$ depends on details of the initial state of the bath spins; {the initial condition for the central spin is such that it} always starts in the ``down'' state.} 

{We first discuss the case of an unpolarized (or minimally polarized) and unentangled bath state; that is, the bath contains $M-1 \approx N/2 -1$ down spins and $N-M$ up spins in a product state. We find that} the Fourier spectra of both the spin expectation value $\langle S_0^z\rangle(t)$ and the entanglement entropy $E(t)$ contain $M$ frequencies. However, these frequencies combine such that $\langle S_0^z\rangle(t)$ approaches {\em zero} within a decoherence time $\tau_d\sim 1/\sqrt{N}$.
Accordingly, $E(t)$ reaches its maximum value 1 within a time $\Or(1/\sqrt{N})$. 
{Deviations from those limits are $\Or(1/N)$.} 
 In other words, in this case, the bath provides a perfect screening of the central spin in the thermodynamic limit.

{Next we still consider a bath state with zero or minimal polarization, but with entanglement among the bath spins, namely an eigenstate of the total bath spin operator. In contrast to the previous case, now} 
{both $\langle S_0^z\rangle(t)$ and $E(t)$ oscillate with only one frequency and maximum amplitude. The frequency depends on the total bath spin {quantum number}. Since the initial eigenstate of the bath is a multipartite entangled state, this raises the question 
{to what extent} an initial bath entanglement generally protects the central spin from decohering. We do not answer this question in the present article, but rather leave it as an interesting perspective for future work.} 

We turn to a partially or fully polarized bath.
Then in the time evolution from an unentangled initial bath state {again}  $M$ frequencies are present, which combine such that the dominating ones are $\Or(N-2M)${, that is, at the lower end of the frequency spectrum}. In the thermodynamic limit, $\langle S_0^z\rangle(t)=-1/2+\Or(1/N)$, $E(t)=\Or((\ln N) /N )$, and even/odd effects are $\Or(1/N)$. {The leading-order non-constant terms oscillate increasingly rapidly as $N$ grows, see Fig. 3 below.}

Finally, the partially {polarized bath}
can be initially in {a total spin} eigenstate such that the dynamics of $\langle S_0^z\rangle(t)$ and $E(t)$ are {again} driven by one frequency only, around a mean value and with an amplitude which depend on the total bath spin {quantum number}.

When comparing these results with the behaviour obtained from a mean-field approach, we find similarity only with the evolution from an initial eigenstate of the bath. This is consistent with \cite{coi06} in so far as there, similarities between the quantum and mean-field approaches were observed for an initially entangled bath. However, the evolution from the unpolarized initially unentangled product state has no classical analogue. {This result is worthwile being stated: Although the exact classical solution of the model \refeq{gaudef} is known \cite{yuz05}, the question which initial bath states reproduce this solution on a quantum-mechanical level has not been addressed yet.}   

This article is organized as follows. In the next section, we briefly sketch our approach for calculating $\rho_0(t)$ and present the dynamical evolution of $\langle S_0^z\rangle(t)$ and $E(t)$. We distinguish between an initially unentangled environment, the environment being initially in an eigenstate and the mean-field solution. Technical details of the calculation are deferred to an appendix. The article ends with an outlook on possible further research directions. 

\section{Results} 
For $A_j\equiv 2 \forall j$ and $h=0$, the Hamiltonian can be written in terms of conserved quantities as follows:
\be
H&=& \vec S^2-\vec S_b^2-\vec S_0^2\label{h},
\ee
where the bath spin $\vec S_b=\sum_{j=1}^{N-1}\vec S_j$ commutes with the Hamiltonian as does $\vec S_0$. The total spin is defined as $\vec S=\vec S_b+\vec S_0$. The Hamiltonian has eigenvalues
\be
E&=&\l\{\begin{array}{ll}
    S_b,&S=S_b+1/2\\
    -S_b-1,& S=S_b-1/2
    \end{array}
    \r.\nn,
\ee
with $S_b=(N-1)/2, (N-1)/2-1,\ldots,N/2-M+1/2$. 

Our aim is to calculate $\langle S_0^z(t)\rangle$ for a fixed value of the total magnetization $S^z=N/2-M$ (in this case, the same symbol is used for the operator and its eigenvalue), starting with an initial configuration $|\!\Down\rangle|\ldots\rangle_b$. We consider the two cases where the initial bath state $|\ldots\rangle_b$ is a pure product state (not entangled) and where the bath is entangled initially. Our approach consists in decomposing the initial state
$|\!\Down\rangle|\ldots\rangle_b$ into a sum over energy eigenstates $|\phi\rangle$ with constant magnetization $S^z$, 
\be
|\!\Down\rangle|\ldots\rangle_b=\sum_\phi c_\phi |\phi\rangle \nn
\ee
where $\phi$ stands for the quantum numbers $S_b, S=S_b\pm 1/2$. 
Starting from this decomposition, the density matrix $\rho(t)$ is written down and therefrom the 
reduced density matrix $\rho_0(t):=\tr_b\l\{\rho(t)\r\}$ of the central spin is calculated, where the bath degree of freedoms are traced {out}.
For any further details on the calculation of $\rho_0(t)$ the reader is referred to appendix \ref{appa}.  
\subsection{Initial product state}
For {an} initial product state $|\!\Down;\down,\ldots,\down,\up,\ldots,\up\rangle$ with $N$ particles, $M_b$ down spins in the bath, $M=M_b+1$ down spins in total, $2M\leq N$, the reduced density matrix for the central spin is calculated in appendix \ref{appa}, Eqs.~(\ref{den1},\ref{den2}). The results for $\rho_0(t)$ and $\langle S_0^z\rangle(t)$ are rewritten here in a slightly different way: 
\be
\fl\rho_0(t)=\sum_{k=0}^{M-1}\frac{(N-M)! (M-1)!}{(N-2k)(N-k)!k!}\l\{2(M-k)(N-M-k)(1-\cos\l[(N-2k)t\r])|\!\Up\rangle\langle\Up|\r.\nn\\
\fl\;\; + \l.\l((M-k)^2+(N-M-k)^2+2(M-k)(N-M-k)\cos\l[(N-2k)t\r]\r)|\!\Down\rangle\langle\Down\!|\r\}\label{rho0}\\
\fl\langle S_0^z\rangle(t)=-\frac12\sum_{k=0}^{M-1}\frac{(N-M)! (M-1)!}{(N-2k)(N-k)!k!}\nn\\
\fl \qquad \times\l\{(N-2M)^2+4(N-M-k)(M-k)\cos\l[(N-2k)t\r]\r\}\label{soz}.
\ee
The entanglement entropy is given by $E(t)=-\tr_0\l\{\rho_0(t)\log_2\rho_0(t)\r\}$. The 
{case $2M > N$ is related to the case $2M \leq N$ by the symmetry property}  
\be
\langle S_0^z\rangle(t)|_M&=&\langle S_0^z\rangle(t)|_{N-M+1}
\quad {(2M>N)}
\label{pol2},
\ee
{in other words, the expectation value of the central spin is invariant under a reversal of all bath spins at $t=0$, with the exception of the case $M=N$, where the system is in a stationary state.}
The evolution from the initial state $|\!\Up;\up,\ldots,\up,\down,\ldots,\down\rangle$ ($N$ particles, $M$ up spins in total) is obtained from Eq.~\refeq{rho0} by exchanging $\Up$ and $\Down$, resulting in an overall sign in Eq.~\refeq{soz}. 
Let us now discuss the results (\ref{rho0}, \ref{soz}) {in more detail}. 

Generally the recurrence time is $\tau_P=\pi$ for even $N$ and $\tau_P=2\pi$ for odd $N$.\footnote{``generally'' means that there are special choices of $N$, $M$ such that $\tau_P$ is shorter than $\pi$. For example, for $M=1$, one has $\tau_P=2\pi \nu/N $, $\nu$ integer.} 
For a fully polarized bath, i.e. $M=1$, only one frequency $N$ is present, and the amplitude of oscillations is $\Or(1/N)$. This case as well as the case $M=2$ have been considered in \cite{kha03}. In the other extreme, for even $N$ and magnetization $S^z=N/2-M=0$, the series {(\ref{soz})} can be summed for $t=\pi/2$, yielding 
\be
\langle S_0^z\rangle(\pi/2)|_{N=2M}&=&\frac{1}{2(N-1)}\sim \frac{1}{2N},\,N\to\infty\label{pi2}\\
E(\pi/2)|_{N=2M}&=&\log_2(N-1)-\frac{N}{2(N-1)}\log_2\frac{N}{2}\qquad\nn\\
& &-\frac{N-2}{2(N-1)}\log_2\l(\frac{N}{2}-1\r)\label{pi2e}\\
&\sim& 1-\frac{1}{2 \ln(2) N^2},\,N\to\infty\nn.
\ee
Applying {Stirling's} formula to the factorials in \refeq{soz} for $N\to \infty$, $M=p N/2$, $0< p <1$, shows that the amplitudes of all but the low-frequency Fourier components are suppressed exponentially. Especially, in the unmagnetized case, {the sum in \refeq{soz} can be evaluated by the saddle point method for large particle numbers and short times, $t\ll \tau_P$. To leading order this yields}
\be
{\langle S_0^z\rangle(t)=-\i |x| \te^{-2 x^2} \cos(2 x \sqrt{N} t) \d x,\; t\ll \tau_P}\label{qsa}.
\ee
{The approximation leading to \refeq{qsa} is commonly called the quasi-static approximation (QSA). Thus in the limit $N\gg 1$ and for times which are short compared to the recurrence time, the particle number only enters through the time scale. We will see below that this time scale plays the role of a decoherence time. }

{We emphasize again that Eq.~\refeq{qsa} describes the time-evolution starting with a pure product state of the whole system. Previous studies \cite{mer02,zha06} concentrated on an initial configuration which involves an average over all bath states and {performed the whole calculation in} the QSA. Our result \refeq{soz}, however, is more general: It is valid without any restrictions{, neither on the particle number nor on the time,} and describes the time evolution from a single product state. As a special case, we may choose the mixed state of \cite{mer02,zha06} as initial state by carrying out the corresponding average. {In this case, we recover the results of \cite{mer02,zha06} in the QSA limit.}}  

\begin{figure}
\begin{center}
\includegraphics*[width=7.5cm]{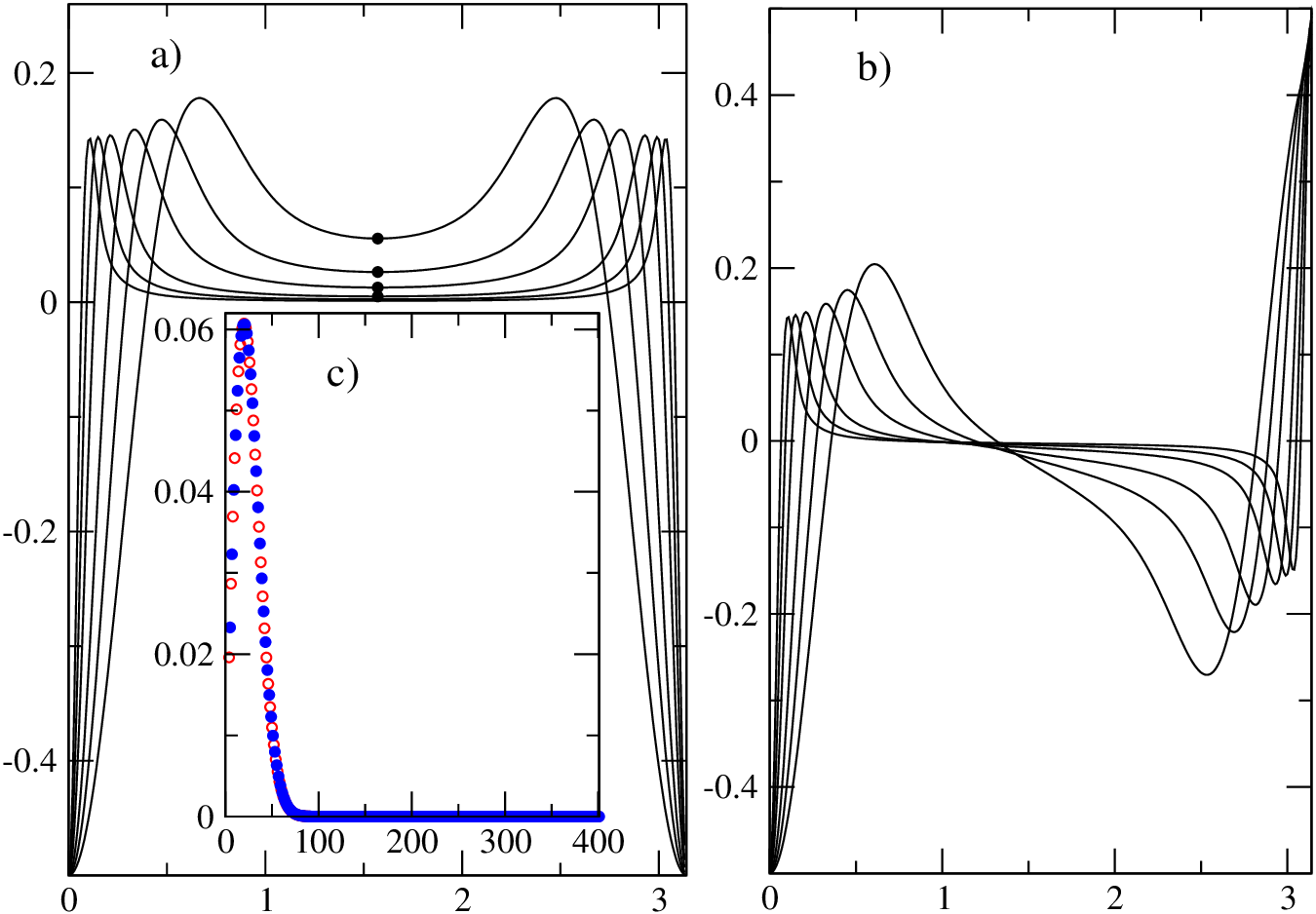}
\caption{{(color online)} Central spin expectation value $\langle S_0^z\rangle(0\leq t\leq \pi)$ in an unpolarized or minimally polarized bath for $N=2M$  in a) and $N=2M+1$ in b), $M=5,10,20,50,100,200$ starting with an unentangled state of the bath. The dots at $t=\pi/2$ in a) are the result \refeq{pi2}. The inset c) shows the Fourier spectrum for $M=200$, $N=400$ (red circles) and $N=401$ (blue dots), as a function of the frequency $N-2k$, $k=0,\ldots,M-1$.} 
\label{fig1}
\end{center}
\end{figure}

{Further} insight is gained by plotting $\langle S_0^z\rangle(t),\,E(t)$ as functions of $t$ for different {values of the bath magnetization}. In Fig.~\ref{fig1}a) the case of even $N=2M$ is depicted. The {function} $\langle S_0^z\rangle(t)$ in the interval $0\leq t\leq \pi$ is symmetric with respect to $t=\pi/2$, showing two zeros and two local maxima. 
The maxima of $\langle S_0^z\rangle(t)$ are found to behave as $\langle S_0^z\rangle(t_{max})=0.14+0.33/N$ with $t_{max}=2.1/\sqrt{N}, \pi-2.1/\sqrt{N}$. The zeros scale like $t_0= 1.3/\sqrt{N},\pi-1.3/\sqrt{N}$.  Consequently, $\langle S_0^z\rangle(\tau_d<t<\pi-\tau_d)\to 0$ for $N=2M\to \infty$, with $\tau_d\sim 1/\sqrt{N}$. Thus $\tau_d$ plays the role of a decoherence time, {as anticipated above}.

\begin{figure}
\begin{center}
\includegraphics*[width=7.5cm]{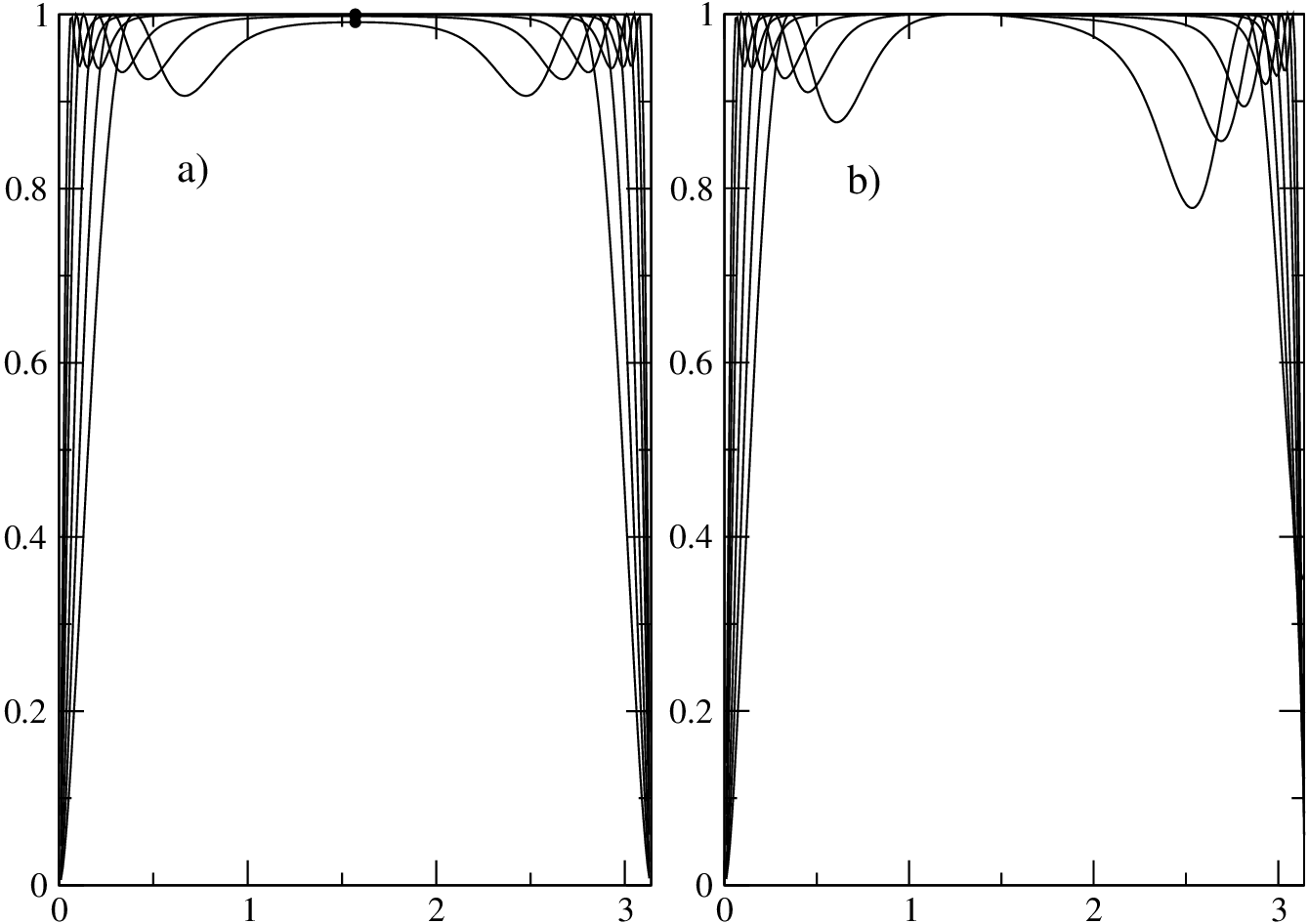}
\caption{Entanglement entropy $E(0\leq t\leq \pi)$ in an unpolarized or minimally polarized bath for $N=2M$ in a) and $N=2M+1$ in b), $M=5,10,20,50,100,200$. The dots in a) at $t=\pi/2$ are the result \refeq{pi2e}. In all cases, the bath is initially unentangled.} 
\label{fig2}
\end{center}
\end{figure}

The entanglement entropy reflects this behaviour: Any deviation of $|\langle S_0^z\rangle(t)|$ from its maximal value $1/2$ results in a finite entanglement entropy of the central spin \cite{sch02}. Especially, a zero in $\langle S_0^z\rangle(t)$ corresponds to a point where $E(t)$ takes its maximal value 1, and a local extremum in $\langle S_0^z\rangle(t)$ translates into a local minimum for $E(t)$, see Fig.~\ref{fig2}a). 
{This means that} the scaling of the zeros and maxima of $\langle S_0^z\rangle(t)$ induces an analogous scaling of the local extrema of $E(t)$, according to Eqs.~(\ref{sden},\ref{eden}).  

{Similar features are} found for the case of odd $N=2M+1$, Fig.~\ref{fig1}b). In this case, however, $\langle S_0^z(t)\rangle$ reaches an overall maximum at $t=\pi$, with $\langle S_0^z(\pi)\rangle\to 1/2$ as $N=2M+1\to \infty$. The width of this maximum (estimated by the zeros next to it) scales $\sim 1/\sqrt{N}$, which means that the {time interval affected by} even/odd effects {(in $N$)} is $\Or(1/\sqrt{N})$. The maximum {$\langle S_0^z \rangle(\pi)$}
 translates into a local minimum  $E(\pi)\to 0 $ with $N=2M+1\to \infty$, cf. Fig.~\ref{fig2}b). 
To conclude our analysis for the case of an unpolarized bath, the Fourier spectrum, i.e. the coefficients  of the oscillating terms in $\langle S_0^z(t)\rangle$, Eq.~\refeq{soz}, as a function of $N-2k$ are plotted in Fig.~\ref{fig1}c). Only low-lying frequencies contribute, high frequencies are suppressed exponentially as observed above.

\begin{figure}
\begin{center}
\includegraphics*[width=7.5cm]{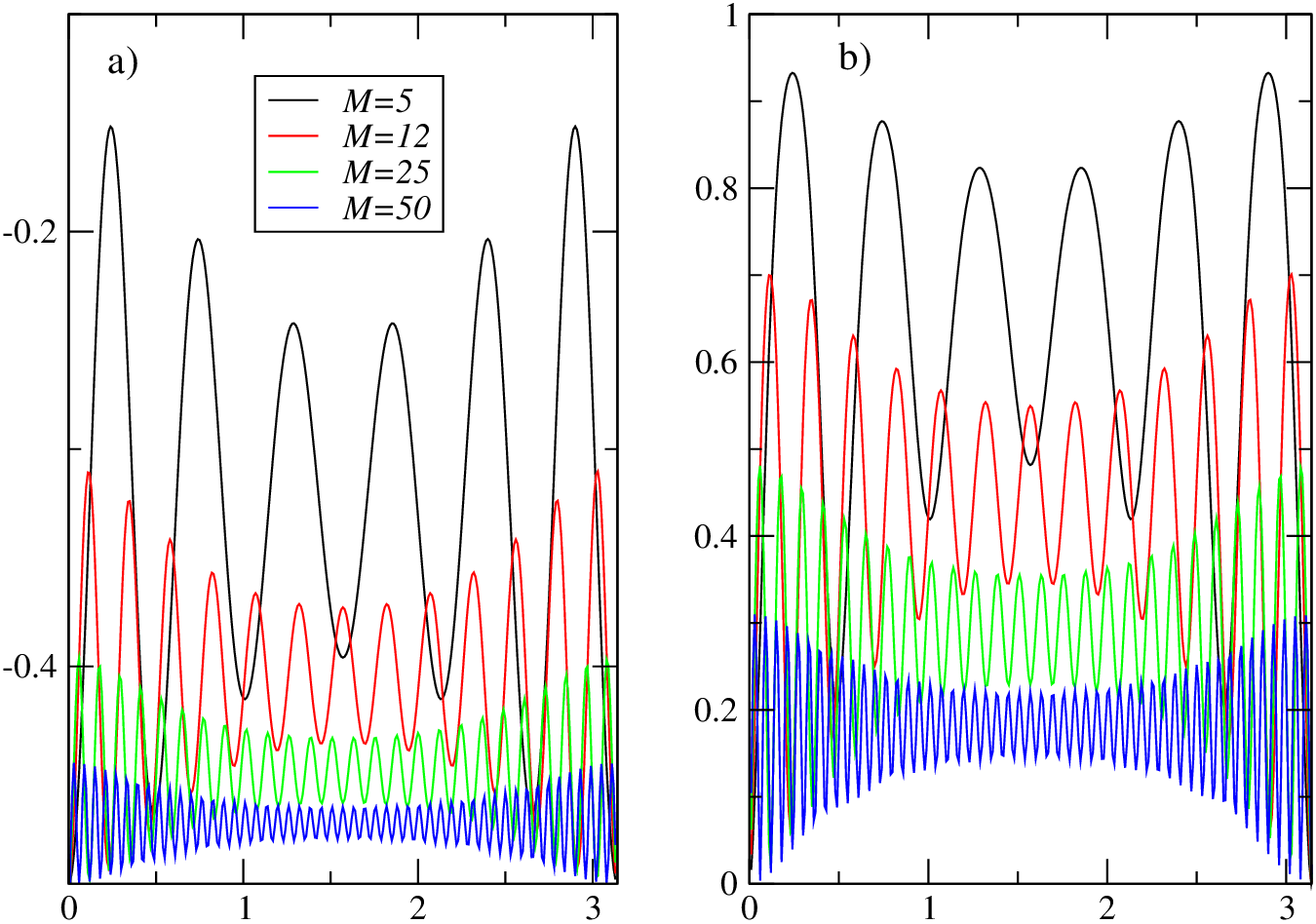}
\caption{{(color online)} Time evolution of a) the central spin expectation value $\langle S^z(t)\rangle$ and b) the entanglement entropy $E(t)$ for partial bath polarization, $N=4M$, $M=5,12,25,50$, respectively. In all cases, the bath state is initially unentangled.} 
\label{fig3}
\end{center}
\end{figure} 

We now consider a partially polarized bath, $M=pN/2$. Again by a numerical finite-size scaling for a variety of choices of $p$, we find that the amplitude of the oscillation {of $\langle S_0^z \rangle(t)$} 
at $t=\pi/2$ is $\Or(1/N)$, around a mean value $-1/2+\Or(1/N)$, independently of $p$. Thus for a partially polarized bath, $\langle S_0^z\rangle(t) = \langle S_0^z\rangle(0)+\Or(1/N)$ as $N\to \infty$. In agreement with this, $E(t)=\Or((\ln N)/N)$. Figs.~\ref{fig3}a), \ref{fig3}b) show $\langle S_0^z\rangle(t)$, $E(t)$ for the special case $N=4M$, or $p=1/2$. 

Note that all these results are valid for homogeneous couplings. The quantum-mechanical time-evolution
{of $\langle S_0^z \rangle(t)$}  for inhomogeneous couplings has been addressed in \cite{kha02,kha03}. In that case, oscillations with frequencies $\Or(N)$, ($\Or(\sqrt{N})$) and amplitudes $\Or(1/N)$, ($\Or(1)$) set in for the initially unentangled polarized (unentangled unpolarized) environment, before the broadening of the couplings leads to a non-exponential decay. That decay at long times does not depend on the particle number. 
{There is no such long-time decay for homogeneous couplings. Instead we observe periodic behaviour with a finite }
recurrence time $\tau_P$ independent of the particle number. For a polarized or partially polarized, initially unentangled bath state, the {oscillation} frequencies and amplitudes are of the same order of magnitude as {they are for inhomogeneous couplings\cite{kha02,kha03} during the initial phase before the long-time decay.}  However, the time evolution starting with an initially unpolarized, unentangled bath as discussed here is governed by a decay after a short time $\tau_d\sim 1/\sqrt{N}$, as opposed to the long-time decay found in \cite{kha02,kha03}. Interestingly, $\tau_d$ {here scales with the particle number in the same way as the period of the initial oscillation in \cite{kha02,kha03} does}.

\subsection{Initial bath eigenstate}
Let the bath initially be in an eigenstate of $\vec S_b^2$ (with eigenvalue $S_b(S_b+1)$) and with magnetization $S_b^z=N/2-M+1/2$. In this state, the bath is entangled initially except for full polarization. 

The reduced density matrix of the central spin for this initially entangled bath state is given in Eqs.~(\ref{rhoe}, \ref{ale}) in appendix \ref{appa}. From these expressions, the time evolution of $\langle S_0^z\rangle$ is deduced
\be
\lefteqn{\langle S_0^z\rangle (t)=-\frac{1}{2(2S_b+1)^2}\l\{(2S_b^z-1)^2 \r.}\nn\\
& & \l.+ 4(S_b-S_b^z+1)(S_b+S_b^z)\cos\l[(2S_b+1)t\r]\r\}\label{sent}.
\ee
The most striking difference to \refeq{soz} is that only one frequency is involved in Eq.~\refeq{sent}. 
Let us consider again the cases of a polarized, partially polarized and unpolarized bath. For a polarized bath one has $S_b^z=S_b=(N-1)/2$. In this case, Eqs.~(\ref{soz},\ref{sent}) yield identical results, which is to be expected because, as mentioned above, a fully polarized bath is in a product state.

For a partially magnetized bath, $S_b^z=p N/2$ with $0<p<1$. This does not fix $S_b$ which can take values $S_b^z\leq S_b\leq (N-1)/2$. {For $S_b=S_b^z$, we obtain}  $\langle S_0^z\rangle (t)=-1/2+\Or(1/N)$, which is similar to the time evolution starting with an initially unentangled state. It is also this behaviour that one might expect from an intuitive point of view: a macroscopic polarization of the bath constitutes an effective gap $\Or(N)$ seen by the central spin, which prevents decoherence. 

However, if {we consider the opposite limiting case when the bath is initially in an} eigenstate with $S_b=(N-1)/2$, then, {in the leading order in $N$}, 
\be
\langle S_0^z\rangle (t)&=&-\frac12\l(p^2+(1-p^2)\cos N t\r)\nn,
\ee
which is an oscillation around the mean value $-p^2/2$ with the frequency $N$, a behaviour qualitatively different from the one at full bath polarization, $S_b=S_b^z$.

Finally, the difference between {an initial product {state} and an initial eigenstate of the} bath becomes most pronounced in the case of small magnetization $\Or(1)$. {Above (see for example Fig. \ref{fig1}) we found that an initial product state of the bath causes the central spin to decohere after a time $\tau_d\sim 1/\sqrt{N}$. However, starting with an initial bath eigenstate,} the time evolution reads {$\langle S_0^z\rangle(t)=-1/2 \cos\l[ (2 S_b+1)t\r]$ for even $N$ and $S_b^z=1/2$.} That is, 
\be
\langle S_0^z\rangle (t) &=&\l\{\begin{array}{cc}
                             -1/2 \cos 2 t,&S_b=1/2\\
                             \vdots  & \vdots \\
                             -1/2 \cos N t,&S_b=(N-1)/2
                             \end{array}\r.\label{qm1},
\ee
whereas for odd $N$ and $S_b^z=0$, {$\langle S_0^z\rangle(t)=-\frac{1}{2(2 S_b+1)^2}\l(1+4 S_b(S_b+1)\cos\l[(2S_b+1)t\r]\r)$}:
\be
\langle S_0^z\rangle (t) &=&\l\{\begin{array}{cc}
                             -1/2 ,&S_b=0\\
                             \vdots  & \vdots \\
                             -\frac{\l(1+N(N-1) \cos N t\r)}{2 N^2},&S_b=(N-1)/2
                             \end{array}\r.\nn.
\ee 
Thus $\langle S_0^z \rangle (t)$ oscillates at maximum amplitude with one of {$\frac{N-1}{2}$} ($\frac N2$) possible frequencies (depending on $S_b$) for odd (even) $N$. {For odd $N$ and $S_b^z=0$, the limiting bath eigenstates are}  the completely symmetric {(maximal $S_b$)} and antisymmetric {(minimal $S_b$)} states. {Note that in the latter case, the complete system is in a zero-energy eigenstate, which means that the time evolution is trivial.} {Especially, for $N=3$, these are the $S^z=0$ triplet and singlet states, respectively, two of the two-qubit Bell states.} 

The behaviour of the entanglement entropy $E(t)$ is found in a straightforward manner from the reduced density matrix in Eqs.~(\ref{rhoe}, \ref{ale}). The qualitative behaviour can be read off from that of $\langle S_0^z\rangle (t)$ by remembering that $\langle S_0^z\rangle=\pm 1/2$ corresponds to $E=0$, $\langle S_0^z\rangle=0$ is equivalent to $E=1$ and any local extrema of $\langle S_0^z\rangle(t)$ are translated into local minima of $E(t)$.

\subsection{Classical/Mean-field solution}
The Gaudin model can be treated in a mean-field approach in the thermodynamic limit \cite{yuz05,coi06}. However, as shown here, when the dynamical evolution of the central spin is considered, the equivalence between the quantum-mechanical and the classical picture does not hold if the initial state is a product state. It does hold, though, if the bath is initially in an eigenstate as {is shown in the following}.    

The mean-field approach consists in replacing the effective field seen by each spin with its quantum-mechanical average \cite{yuz05,coi06}. We denote the corresponding quantities by an additional {subscript} $MF$, 
\be
H\to \l\{\begin{array}{c}
          2 \vec S_0 \cdot\vec S_{b,MF}\\
          2 \vec S_{0,MF}\cdot\vec S_b
          \end{array}\r.\nn\;.
\ee
Since the resulting equations of motion $\dot{\vec S}_{0,b}=\rmi \l[H,\vec S_{0,b}\r]=2\vec S_{0,b}\times \vec S_{b,0,MF}$ are linear in the quantum mechanical operators, expectation values are taken here as well and one arrives at {the classical equations of motion}
\be
\dot{\vec S}_{0,MF}&=&2\vec S_{0,MF}\times \vec S_{b,MF}\nn\\
\dot{\vec S}_{b,MF}&=&2\vec S_{b,MF}\times \vec S_{0,MF}\nn.
\ee
{}From these equations it follows that $2(\vec S_{0,MF}+\vec S_{b,MF}):=\vec g = const.$, as expected. Let us choose the {coordinate system}
such that $\vec g =(0,g,0)^T$, and the initial conditions $\vec S_{0,MF}(0)=(0,0,S_0)^T$, $\vec S_{b,MF}(0)=(0,g/2,-S_0)^T$. Then 
\be
\vec S_{0,MF}(t)=\l(\begin{array}{c} 
                  -S_0 \sin g t\\
                  0\\
                  S_0 \cos g t \end{array}\r)\nn.
\ee
Comparing this result with the quantum-mechanical solution starting from an initial bath eigenstate {\refeq{qm1}}, one notices that both results are closely related, {where the parameter $g/2$ in the classical solution plays the role of $S_b$ in the quantum-mechanical picture.} Note that in both cases, the polarization of the bath along the $z$-axis is of the same order as the central spin. However, for an initially unentangled bath state, the {quantum-mechanical solution does not match the mean-field behaviour}. 

{This leads to the question {to what extent} initial entanglement in the bath is necessary in order to reproduce the classical solution within the quantum-mechanical framework. In \cite{coi06} it was observed that if  a spin-coherent (i.e. entangled) state is chosen as the initial bath state, the quantum-mechanical solution is recovered by the mean-field solution. Given that the exact classical solution of the model \refeq{gaudef} has been obtained recently \cite{yuz05}, addressing that question quantitatively remains an interesting task for the future}. 

\section{Conclusion and outlook} 
We {have} studied the time evolution of the central spin in the central-spin model with homogeneous couplings and without magnetic field by calculating the corresponding reduced density matrix. Whereas the mean-field solution is an appropriate description for the evolution starting with the bath being unpolarized and initially in an eigenstate, the full quantum mechanical solution is needed for an initially unentangled unpolarized bath. In this case, the central spin is completely screened by the bath spins after a decoherence time $\tau_d\sim 1/\sqrt{N}$ until $\pi-\tau_d$, within deviations $\sim 1/N$ for a non-polarized bath in the thermodynamic limit. {The {central-spin polarization recurs after the} recurrence time $\tau_P=\pi (2\pi)$ for $N$ even (odd). A complete {polarization reversal} at $\tau_P/2=\pi$ occurs for $N$ odd; however, the {duration of this reversal} is $2 \tau_d$, so that in this respect, even/odd-effects are $\Or(1/\sqrt{N})$}. The dynamics of the entan
 glement entropy of the central spin reflects the dynamics of the central spin: For an unpolarized initial product state, the central spin finds itself {maximally entangled with the bath spins} after a time $\Or(1/\sqrt{N})$. Thus for an initially unentangled bath, non-trivial dynamics are obtained only from the full quantum-mechanical solution at finite particle number - in the thermodynamic limit {the decoherence time vanishes, and after that}, the dynamics is frozen. 

Directions for future research include the investigation of the effect of {initial entanglement within the bath}. Two questions are of particular importance in this context: 
(i) Whether it is generally true that initial bath entanglement protects the central spin from decoherence (the eigenstates of a non-polarized bath considered in section II.B are entangled - but at this stage it is unclear whether it is rather the eigenstate- or the entanglement-property that is important). {A hint to a positive answer to this question can be found in \cite{daw05}, where it was argued that persistent entanglement in the bath serves as a barrier against decoherence of the central spin. The second question} (ii) is under which conditions initial bath entanglement allows to recover the mean-field solution. 
Furthermore, the exact generalization of the quantum mechanical solution to inhomogeneous couplings is highly desirable. First results have been obtained \cite{tbp} in the form of an exact formula for $\langle S_0^z\rangle(t)$ stemming from the Bethe Ansatz. The explicit evaluation of this formula {especially for a non-polarized bath} is the subject of future work. 

\vspace{-0.2cm}
\section*{Acknowledgments}
{We thank} M.T. Batchelor and X.-W. Guan for helpful discussions. This work has been supported by the German Research Council (DFG) under grant number BO2538/1-1.
\vspace{0.5cm}
\appendix
\section{Calculation of the reduced density matrix}
\label{appa}
We first have to decompose the initial state {$$|\!\Down;\down,\ldots,\down,\up,\ldots,\up\rangle=:\\|\!\Down\rangle|\!\down,\ldots,\down,\up,\ldots,\up\rangle_{N-1}$$}with $N$ particles and $M=M_b+1$  down-spins, $2M\leq N$, into eigenstates of $H$.

For a given value of the bath spin $S_b$, there are two values of the total spin, namely $S=S_b+1/2$ and $S=S_b-1/2$. We denote an $N$-particle eigenstate of $\vec S^2$ (with eigenvalue $S(S+1)$), {$\vec S_b^2$ (with eigenvalue $S_b(S_b+1)$)} and $S^z$ (with eigenvalue $S^z$) as $|S,{S_b,}S^z\rangle_N$. {Furthermore, an $N$-particle eigenstate of $\vec S^2$ and $S^z$ is denoted by $|S,S^z\rangle_N$.} The total angular momentum eigenstates of an arbitrary angular momentum plus a spin 1/2 are a standard example (cf. for example \cite{schwabl}) in quantum mechanics texts. For the present case one obtains that \cite{schwabl} 
\be
\fl\l|S_b+\frac12,{S_b},S_b^z-\frac12\r\rangle_N&=&\frac{1}{\sqrt{2S_b+1}} \l[\sqrt{S_b+S_b^z}|\!\Up\rangle|S_b,S_b^z-1\rangle_{N-1}\r.\nn\\
& &\l.+\sqrt{S_b-S_b^z+1}|\!\Down\rangle|S_b,S_b^z\rangle_{N-1}\r]\label{es1}\\
\fl\l|S_b-\frac12,{S_b},S_b^z-\frac12\r\rangle_N&=&\frac{1}{\sqrt{2S_b+1}}\l[\sqrt{S_b-S_b^z+1}|\!\Up\rangle|S_b,S_b^z-1\rangle_{N-1}\r.\nn\\
& &\l.-\sqrt{S_b+S_b^z}|\!\Down\rangle|S_b,S_b^z\rangle_{N-1}\r]\label{es2}
\ee
are eigenstates of $\vec S_b^2$ with eigenvalue $S_b(S_b+1)$ and of $H$ as well with eigenvalues $S_b$, $-S_b-1$, respectively. {Note that the states on the right-hand side of these equations are product states between the bath (which is specified by its total spin and the total spin's $z$-component) and the central spin.} {The eigenstates (\ref{es1},\ref{es2}) {were also} constructed in \cite{rya82}, where the partition function of the model was studied}. The relations (\ref{es1},\ref{es2}) can be inverted to obtain 
\be
\fl\l|\!\Down\r\rangle\l|S_b,S_b^z\r\rangle_{N-1}&=&\frac{1}{\sqrt{2S_b+1}} \l[\sqrt{S_b-S_b^z+1}\l|S_b+\frac12,{S_b},S_b^z-\frac12\r\rangle_N\r.\nn\\
& &-\l.\sqrt{S_b+S_b^z}\l|S_b-\frac12,{S_b},S_b^z-\frac12\r\rangle_N\r]\label{first}\\
\fl\l|\!\Up\r\rangle\l|S_b,S_b^z-1\r\rangle_{N-1}&=&\frac{1}{\sqrt{2S_b+1}}\l[\sqrt{S_b+S_b^z}\l|S_b+\frac12,{S_b},S_b^z-\frac12\r\rangle_N\r.\nn\\
& &+\l.\sqrt{S_b-S_b^z+1}\l|S_b-\frac12,{S_b},S_b^z-\frac12\r\rangle_N\r]\label{sec}.\ee
If the initial state of the bath is an eigenstate of $\vec S_b^2$ and $S_b^z$ (and hence an entangled state in all but the fully polarized cases),
 \refeq{first} can be used to construct the  time-dependent reduced density matrix $\rho_0(t)$ for the central spin. 
The projection operator $|\Phi(t)\rangle \langle \Phi(t)|$ developing from the initial state of the full system contains combinations of the Hamiltonian eigenstates (\ref{es1},\ref{es2}). {From that projection operator, $\rho_0(t)$ is obtained as $\rho_0(t)=\tr_b\l\{|\Phi(t)\rangle \langle \Phi(t)|\r\}$.} Thus the trace over the bath spins for each of {those} combinations {has to be performed};
for example (\ref{es1}) yields
\be
\tr_b \l\{\l|S_b+\frac12,{S_b},S_b^z-\frac12\r\rangle_{N}\, \l\langle S_b+\frac12,{S_b},S_b^z-\frac12\r|_N\r\}\;\;\nn\\
 \qquad =\frac{1}{2S_b+1} \l[(S_b+S_b^z)|\!\Up\rangle\langle\Up|+(S_b-S_b^z+1)|\!\Down\rangle\langle\Down\!|\r]\nn.
\ee
and similar expressions for the other combinations. The final result is
\be
\rho_0(t)&=&\alpha(t)|\!\Up\rangle\langle\Up|+(1-\al(t))|\!\Down\rangle\langle\Down\!|\label{rhoe}
\ee
\be
\al(t)&=&\frac{2(S_b-S_b^z+1)(S_b+S_b^z)}{(2S_b+1)^2}\nn\\
& & \times (1-\cos[(2S_b+1)t]), \label{ale}
\ee
{implying that for an initial eigenstate of the total bath spin all physical quantities oscillate harmonically with frequency $2S_b+1$.}
In order to find $\rho_0(t)$ for the case of an initially unentangled bath state, we proceed as follows: 
Starting from the completely polarized state $|\!\up,\ldots,\up\rangle_{N-M}$,  the decomposition of $|\!\Down\rangle |\!\up,\ldots,\up\rangle_{N-M}$ can be found as above. We then consider this state as the new bath state $|\!\down,\up,\ldots,\up\rangle_{N-M+1}$ and find the decomposition of $|\!\Down\rangle|\!\down,\up,\ldots,\up\rangle_{N-M+1}$. This procedure is repeated until the state $|\!\Down\rangle|\!\down,\ldots,\down,\up,\ldots,\up\rangle_{N-1}$ is reached. The result is of the following form
\be
\fl|\!\Down\rangle|\!\down,\ldots,\down,\up,\ldots,\up\rangle_{N-1}=\sum_{k=0}^{M-1}\frac{1}{\sqrt{N-2k}}\l(\sum_{j=1}^{C^{M-1}_k}c_{j,k}^{(N,M)}\r)\nn\\
\fl\qquad \times \l(\sqrt{M-k}\l|\frac{N}{2}-k,\frac{N-1}{2}-k{,\frac{N}{2}-M}\r\rangle_N\r.\nn\\
\fl \qquad \l.-\sqrt{N-M-k}\l|\frac{N}{2}-k-1,\frac{N-1}{2}-k{,\frac{N}{2}-M}\r\rangle_N\r)\label{prod},
\ee
with unknown real coefficients $c_{j,k}^{(N,M)}$. The symbol $C^N_M$ denotes the binomial coefficient $N!/(M!(N-M)!)$. Note that the 
total number of states is $2\cdot \sum_{k=0}^{M-1}C^{M-1}_k=2^M$, as expected. All states are mutually orthogonal, so that in the expressions for the density matrix, the sum of squares $d_{k,N,M}:=\sum_{j=1}^{C^{M-1}_k}\l[c_{j,k}^{(N,M)}\r]^2$ enters and it is this quantity that we set up a recursion relation for. 
The time-dependent reduced density matrix $\rho_0(t)$ corresponding to the state (\ref{prod}) is
\be
\rho_0(t)&=& \beta(t)|\!\Up\rangle\langle\Up|+(1-\beta(t))|\!\Down\rangle\langle\Down\!|\label{den1}
\ee
\be
\beta(t)&=&2\sum_{k=0}^M \frac{d_{k,N,M}}{(N-2k)^2}(M-k)(N-M-k)\nn\\
& & \times(1-\cos[(N-2k)t])\label{den2},
\ee
from which we derive
\be
\langle S_0^z(t)\rangle&=&-\frac{1}{2}(1-2\beta(t))\label{sden}\\
E(t)&=& -\beta(t)\log_2\beta(t)\nn\\
& & -(1-\beta(t))\log_2\l[1-\beta(t)\r]\label{eden}.
\ee
{Hence, for an initial product state of the bath (with $M-1$ reversed bath spins) a spectrum of $M+1$ frequencies  $N-2k$ $(k=0, \ldots, M)$ governs the dynamics. In order to further determine the function $\beta(t)$ (\ref{den2}), we next}
decompose the $(N-1)$-spin bath part of the state (\ref{prod}) into a single down spin and a $(N-2)$-spin state with $(M-1)$ down spins in a way completely analogous to (\ref{prod}), with the replacements
$N\to N-1$, $M\to M-1$, and calculate $|\!\Down\rangle|\!\down,\ldots,\down,\up,\ldots,\up\rangle_{N-1}$ again. 
By comparing coefficients in $\langle S_0^z(t)\rangle$, the following recursion relation is found for $d_{k,N,M}$:
\be
d_{k,N,M}&=& \frac{M-1-k}{N-1-2k}d_{k,N-1,M-1}\nn\\
& &+\frac{N-M-(k-1)}{N-1-2(k-1)} d_{k-1,N-1,M-1}\nn 
\ee
with $d_{0,N,M}=\l[C^{N-1}_{M-1}\r]^{-1}$. 
This recursion relation is solved by
\be
d_{k,N,M}&=& \frac{(M-1)!(N-M)!(N-2k)}{(N-k)!k!}\nn
\ee
which leads to the result Eq.~\refeq{rho0}. Note that $\sum_{k=0}^{M-1}d_{k,N,M}=1$, which is the normalization condition for the state Eq.~\refeq{prod} and at the same time guarantees $\langle S_0^z\rangle(0)=-1/2$. 

The calculation for $2M>N$ {proceeds} analogously. Now the state $|\Down\rangle|\up,\ldots,\up,\down,\ldots,\down\rangle_{N-1}$ with $N-M$  $\up$-spins is built successively starting from the polarized state $|\down,\ldots,\down\rangle_{M-1}$. In each step, one $\up$-spin is added according to Eq.~\refeq{sec}, until finally, $|\Down\rangle$ is included using Eq.~\refeq{first}. {It tuns out that in this case the dynamics of the central spin is the same as in the case of $N+1-M$ down spins and $M-1$ up spins in the bath. In other words, $\langle S_0^z \rangle (t)$ is invariant under reversal of all bath spins except for the case of a fully polarized bath, where reversal of all bath spins leads to a stationary state of the entire system. This leads to Eq.~\refeq{pol2} in the main text.}    

\section*{Bibliography} 
\bibliography{gaudin} 
\end{document}